\documentstyle[epsfig,epsf,aps]{revtex}
\begin{document}
\draft
\title{Centrifugally driven relativistic dynamics on curved trajectories}
\author{Andria Rogava}
\address{Dipartimento di Fisica Generale, Universit\'a degli Studi di Torino,
Via Pietro Giuria 1, I-10125 Torino; and
Abdus Salam International Centre for Theoretical Physics, I-34014 Trieste,
Italy}
\author{George Dalakishvili}
\address{Department of Physics, Tbilisi State University,
Chavchavadze ave. 2, Tbilisi 380028, Georgia}
\author{Zaza Osmanov}
\address{Centre for Plasma Astrophysics, Abastumani  Astrophysical
Observatory, Kazbegi str. $2â$, Tbilisi 380060, Georgia}
\date{\today}
\maketitle

\begin{abstract}

Motion of test particles along  rotating curved trajectories is
considered. The problem is studied  both in the laboratory and the
rotating frames of reference. It is assumed that the system
rotates with the constant angular velocity $\omega = const$. The
solutions are found and analyzed for the case when the form of the
trajectory is given by an Archimedes spiral. It is found that
particles can reach infinity while they move along these
trajectories and the physical interpretation of their behaviour is
given.
The analogy of this idealized study with the motion of particles
along the curved rotating magnetic field lines in the pulsar
magnetosphere is pointed out. We discuss further physical
development (the conserved total energy case, when $\omega \ne
const$) and astrophysical applications (the acceleration of
particles in active galactic nuclei) of this theory.

\end{abstract}


\section{Introduction}

Rotation and relativity are those two features of motion, which do
{\it not} easily match with each other. Still in astrophysics,
with its abundance of extremely strong electromagnetic and
gravitational fields, there are situations where motion {\it is}
both rotational and relativistic. Most prominent examples include
swirling astrophysical jets in active galactic nuclei (AGNs) and
quasars, innermost regions of black hole accretion disks,
accretion columns in X-ray pulsars and plasma outflows in radio
pulsar magnetospheres. In these {\it kinematically complex
astrophysical flows}, where rotation is interlaced with the
relativistic motion of particles, the coexistence of these two
features of the motion leads to observationally puzzling phenomena
with sophisticated and ill-understood physical background. The
interest to these flows is not new, but the upgrade of highly
idealized models to more realistic, astrophysically relevant
levels is still related with major theoretical and computational
difficulties.

Some important and basic theoretical issues, related with the
relativistic rotation, are not uniquely defined and often evoke
controversial interpretations. One of the
most notable examples is the {\it ``centrifugal force reversal"}
effect, originally found in [1], and later [2-4] studied in detail. It was
argued that under certain conditions the centrifugal
force {\it attracts} towards the rotation axis both for
Schwarzschild and Kerr black holes. In Ernst spacetime, which
represents the gravitational field of a mass embedded in a
magnetic field, the centrifugal force acting on a particle in
circular orbit was reported [5] to reverse its sign even twice! In
the simplest case of the Schwarzschild spacetime, strictly  and
essentially speaking, it was found that below the radius of the
spatially circular photon orbit an {\it increase} of the angular
velocity of a test particle causes more {\it attraction} rather than
additional centrifugal {\it repulsion}. This effect  was interpreted by
Abramowicz [4] in terms of the centrifugal force reversal - it was
stated that in such cases the centrifugal force {\it attracts}
towards the axis of rotation!

This interpretation was criticized by de Felice [6,7] (see also
[8]), who argued that the discovered effect could be attributed to
the strength of the gravitational field and be explained in a way
which preserves the repulsive character of the centrifugal force.
The spirit of this approach --- to save the intuitively appealing
nature of the centrifugal force as of {\it ``something which
pushes things away"} [7] --- is theoretically valid and
practically convenient. After all, in general relativity, there is
no implicit way to define the centrifugal force: in any case one
needs to introduce some sort of ``3+1" spacetime splitting and dub
as the ``centrifugal force" some Newtonian-like expression, which
looks like it [7]. Moreover, de Felice found several interesting
examples of the ambiguity of the global concept of 'outwards´ and
pointed out at the deep interrelation of this problem with
the definition of the centrifugal force in relativity. Abramowicz
studied further the problem of the local and the global meaning of
'inwards´ and 'outwards´ [9] and showed that the centrifugal force
always repels outwards in the {\it local} sense, while it may
attract inwards, towards the centre of the circular motion, in the
{\it global} sense! The theoretical scheme for the operationally
unambiguous definition of the inward direction was suggested by de
Felice and Usseglio-Tomasset [10] and later this approach was used
for the geometrical definition of the generalized centrifugal
force [11].

Therefore the effect, discovered in [1], is indubitably a genuine
relativistic effect, although its {\it interpretation} in terms of the
``reversal" of the centrifugal force is not implicit and is
largely the matter of definition.

Same is true for another rotational effect, disclosed by Machabeli
and Rogava [12], on the basis of the relatively simple and
idealized special- relativistic {\it gedanken experiment}: motion
of a bead within a rigidly and uniformly rotating massless linear
pipe. It was shown that even if the starting velocity of the bead
is nonrelativistic, after an initial phase of usual centrifugal
acceleration, while the bead acquires high enough relativistic
velocity, it starts to {\it decelerate} and after reaching the
light cylinder changes the character of its motion  from
centrifugal to centripetal. It was found that when the initial
velocity $v \ge \sqrt{2}/2$ the motion of the bead is
decelerative all the way from the pivot to the light cylinder.

Certainly no real pipe may stay absolutely rigid, especially
nearby the light cylinder. Besides, in order to maintain the
uniform rotation of such a device, one needs an infinite amount of
energy. Therefore the setup considered in [12] was {\it highly
idealized}. The constant rotation rate assumption was replaced in
[13] with a more realistic one: the total energy of the
system ``rotator+pipe+bead" was assumed to be constant. It was found
that the moving bead acquires energy from the slowing down
rotator, but under favorable conditions the bead deceleration
still happens.

The results of [12] were interpreted by its authors in terms of
the centrifugal force reversal. De Felice disputed this
interpretation [14] and argued that, also in this case, like in
above-mentioned general-relativistic examples, the generalized
definition of the centrifugal force may guarantee the absolutely
repulsive character of the force. He  pointed out
that an inertial observer will never {\it see} the bead reaching
the light cylinder, because all light signals from the bead are
infinitely redshifted. It was also shown that the vanishing of the
radial velocity of the bead at the light cylinder can be interpreted
in terms of the corresponding vanishing of the bead's proper time.

Despite the controversy of interpretations it is generally believed
that rotational relativistic
effects could operate in different astrophysical situations and
might, hopefully, lead to detectable observational appearances.
Recently, Heyl [15] suggested that the observed QPO frequency
shifts in bursters are caused by a geometrical effect of the
strong gravity, similar to the Abramowicz-Lasota centrifugal force
``reversal"\footnote{However later [16] it was found that the Heyl's
calculations contained the sign error and the real effect could
hardly account for the observed frequency shifts in the type I
X-ray bursts.}. As regards Machabeli-Rogava {\it gedanken} experiment,
it implies that radially constrained, relativistic and rotationally
(centrifugally) driven motion shows inevitable radial deceleration near
the light cylinder. Evidently this effect might occur in a number of
astrophysical situations, where motion is constrained, rotational
and strongly (special) relativistic.

One of the most important class of astrophysical flows, where
this effect could show up, is {\it centrifugally driven
outflows}. In the context of pulsar emission theory they were
first considered in the late 1960s by Gold [17,18] (for recent
studies see e.g. Ref.19 and 20). For accreting black holes, both
of galactic and extragalactic origin, Blandford and Payne [21]
first noted that centrifugally driven outflows from accretion
disks could be responsible for the launch of jets, if the poloidal
field direction is inclined at an angle less than $60^\circ$ to
the radial direction\footnote{For a rapidly rotating Kerr black
hole the critical angle can be as large as $90^\circ$ [22].}.

Recently Gangadhara and Lesch [23] suggested that centrifugal
acceleration, taking place as a consequence of the
bead-on-the-wire motion similar to the Machabeli-Rogava [12] {\it
gedanken} experiment, may account for the acceleration of
particles to very high energies by the centrifugal forces while
they move along rotating magnetic filed lines of the rotating AGN
magnetosphere. They claimed that the highly nonthermal, X-ray and
$\gamma$-ray emission in AGNs arises via the Comptonization
(inverse-Compton scattering) of ultraviolet photons by
centrifugally accelerated electrons. The same processes was
critically re-examined by Rieger and Mannheim [24] and it was
found that the rotational energy gain of charged particles is
efficient but substantially limited not only by the Comptonization
but also by the effects of the relativistic Coriolis forces. The
specific nature of the propagation of electromagnetic radiation in
the rotating frame of reference [14,25] is another aspect of this
problem, which still needs to be taken into account.

The whole philosophy of the `pipe-bead' (ot `rotator-pipe-bead')
{\it gedanken} experiments was to mimic the common situation in
relativistic and
rotating astrophysical flows, where the plasma particles are
doomed to move along the field lines of governing magnetic fields.
While we consider relatively small length scales, the shape of the
field lines can be approximated as being straight.  However, on
larger length scales the curvature of the field lines turns out to
be important for the physics of the plasma streams, which are
guided by them. The natural question arises: how the motion of the
bead changes when the pipe is curved? In other words, how the
dynamics of particles, prescribed to move along the fixed
trajectories, change when the shape of their involuntary tracks of
motion is {\it not} straight!? Obviously, this is not only a mere
theoretical curiosity, but the issue which might have a tangible
practical importance. In astrophysical situations the role of the
``pipes" is played by the magnetic field lines, and the latter are
{\it always} curved. Therefore, it is clear that the study of the
motion of test particles along prescribed curved rotating
trajectories is a necessary and important step for the ultimate
building of a physically meaningful model of centrifugally driven
relativistic particle dynamics for rotating magnetospheres of
pulsars and AGNs.

It is the purpose of this paper to address the above stated issue.
In particular, in the next section, we develop
special-relativistic theory of the motion of centrifugally driven
particles on fixed nonstraight trajectories. The formalism is
developed both for the laboratory frame (LF) and for the frame of
reference rotating with the system (rotating frame, or RF).
Equations of motion are derived and solved numerically. The
detailed study is given only for the case when the angular
velocity of the rotation is constant. However, we also outline the
formalism for the astrophysically more realistic case of the
conservative `rotator-pipe-bead' system with perceptible exchange of
energy between the bead and the rotator, leading to the
variability of the angular velocity of the whole system. In the
final section of the paper we discuss the results, consider the
directions and aims of the future study, suggest and discuss those
astrophysical situations,  where the obtained results could be
useful for the clarification of puzzling observational appearances
of related astronomical objects.

\section{Main consideration}

The ideal two-dimensional system, which we are going to consider,
consists of three basic parts: the device of the mass $M$ and the
moment of inertia ${\cal I}$, rotating with the angular velocity
$\omega(t)$, hereafter referred as the {\it rotator}; the massless
but absolutely rigid {\it pipe} steadily attached to the rotator;
and the  small {\it bead} of the mass $m$ and the  radius equal to
the internal cross-section radius of the pipe. The bead is put
inside the pipe and can slide along the pipe without a friction.
Evidently, instead of the pipe-bead dichotomy, since we are
considering the two-dimensional layout, one may think about the
'wire-on-bead´ analogy, which is sometimes used [24].

Contrary to the [12], where a straight pipe case was studied, now
we let the pipe to be an arbitrarily flat curve, mathematically defined by:
$$
\varphi\equiv
\varphi(r), \eqno(1a)
$$
with $[d\varphi/dr \equiv{\varphi}'(r)]$:
$$
\Phi \equiv r{\varphi}'(r). \eqno(1b)
$$

The dynamics of the system may be studied basing on two alternative
assumptions:

\begin{enumerate}

\item
It makes the task simpler to suppose that the kinetic  energy of
the rotator $E_M$ is huge and always $E_M \gg E_m$; i.e., despite the
exchange of the energy
with the moving bead, $E_M$ stays practically constant. Hence, the
angular velocity of the whole ``rotator-pipe-bead" system
(henceforth referred as the RPB system) stays constant:
$$
\omega = const. \eqno(2a)
$$

In this case the first part of the triple RPB system (the rotator)
continuously supplies the bead with energy and helps to keep the
angular velocity of rotation constant. Therefore, the problem
reduces to the study of the pipe-bead double system (the PB
system) with the {\it constant} rotation rate. In the case of the
{\it straight} pipe ($\varphi=\varphi_0$) the problem has exact
analytic solution, found and analyzed in [12].

\item

It is more realistic to assume that the rotator energy $E_M$ is
finite, so that the whole RPB system is conservative
$E_{tot}\equiv E_M+E_m=const$. There is a perceptible energy
exchange between the rotator and the bead: both the energy $E_M$
and the angular momentum $L_M$ of the rotator {\it are} variable
and, consequently, the angular velocity of the rotation can {\it
not} stay constant:
$$
\omega \ne const. \eqno(2b)
$$
The problem with the straight pipe and variable rotation rate $\omega(t)$ has
no analytic solution. It was studied numerically in [13].

\end{enumerate}

With either (2a) or (2b) assumptions the pipe is always assumed to be
the passive part of the system. In order to mimic a magnetic field
line it is assumed to be massless, having no share in
the energy and/or momentum balance of the whole system.  Still the role of
the pipe --- as the dynamic link between the rotator and the
bead --- is significant: it provides the prescribed ``guiding" of the bead motion
in the rotating frame of reference and makes the trajectory of the bead
known in advance.

In this paper the dynamics of the {\it gedanken} system is studied in detail
only under the first, easier, assumption of the constant angular velocity.
The rout to the solution of the problem under the second
assumption is also given, but its full study needs separate
consideration and will be published elsewhere.

There are two natural frames of reference, in which the dynamics
of this system could be studied. The first, inertial one, is the
laboratory frame (LF), where the observer measures the angular
velocity of the rotator (and the pipe) to be $\omega(t)$, while the angular
velocity of the bead is equal to:
$$
\Omega(t)=\omega(t)+{\varphi}'(r)v(t), \eqno(3)
$$
and the dynamics of the moving bead is governed by the pipe reaction
force acting on it. Note that $v(t) \equiv dr/dt$  is the radial velocity of
the bead relative to the LF.

The second frame, rigidly attached to the rotator and rotating
with it (hereafter referred as the rotating frame, or the RF), is
non-inertial, but quite convenient for the inspection of the motion of the
bead along the curved pipe.
This original approach implies embodying of the form of the pipe  into the
metric of the rotating frame. It was
used in [12] for the straight pipe case and proved to be quite
efficient for the case when $\omega(t)$ is assumed to be a
constant. On the contrary, as we shall see later, the LF treatment appears
to be handier when the second (2b) approximation ($E_{tot}=const$ and
$L_{tot}=const$, rather than $\omega(t)=const$) is chosen. That is why it is
important to consider the problem both in the LF and in the RF.

\subsection{Uniformly rotating PB system}

First, let us consider the problem in the laboratory frame of
reference (LF) and ascertain that it admits full (numerical) solution of
the associated initial value problem. Second, let us consider the
same problem in the rotating frame of reference (RF). We shall see
that when the (2a) assumption of the constancy of the rotation rate is used
the latter approach is mathematically easier and
provides fuller information about the dynamics of the system.

\subsubsection{LF treatment}

The most straightforward way to approach the problem is to consider
it in the laboratory frame of reference, in which the spacetime is
Minkowskian:
$$
ds^2 =- dT^2 + dX^2 + dY^2=- dT^2 + dr^2 + r^2d\phi^2. \eqno(4)
$$

We use geometrical units, in which $G=c=1$. Note that azimuthal
angle $\phi$, as measured in the LF, is related with the azimuthal
angle $\varphi$, measured in the RF, via the obvious expression:
$\phi = \varphi + \omega t$. The pipe reaction force ${\bf F}$  is
the dynamic factor constraining the bead to move along the pipe. It
is easy to see (from the 4-velocity normalization
$g_{\alpha\beta}U^{\alpha}U^{\beta}=-1$) that the Lorentz factor
of the moving bead is:
$$
\gamma(t)=[1-r^2\Omega^2-v^2]^{-1/2}. \eqno(5)
$$

The angle between the radius-vector of a point of the pipe and
the tangent to the same point is given by the relation:
$$
\alpha = \arctan \Phi, \eqno(6)
$$
and the components of the reaction force, acting in the radial and
azimuthal directions, are
$$
F_r = -|F|\sin{\alpha} = -{{\Phi}\over{\sqrt{1+\Phi^2}}}|{\bf F}|,
\eqno(7a)
$$
$$
F_{\phi} = |F|\cos{\alpha} = {{1}\over{\sqrt{1+\Phi^2}}}|{\bf F}|,
\eqno(7b)
$$
respectively.

Defining the physical components of the bead relativistic momentum
$[m(t) \equiv m_0\gamma(t)]$:
$$
P_r \equiv  mv, \eqno(8a)
$$
$$
P_{\phi} \equiv mr{\Omega}, \eqno(8b)
$$
we can write the two components of the equation of motion in the
following way:
$$
\dot{P}_r - {\Omega}P_{\phi}=F_r, \eqno(9a)
$$
$$
\dot{P}_{\phi} + {\Omega}P_r=F_{\phi}. \eqno(9b)
$$

Combining these equations we can, first,
derive the equation:
$$
\dot{P}_r+{\Phi}\dot{P}_{\phi}+{\Omega}
({\Phi}P_r-P_{\phi})=0. \eqno(10a)
$$
It is easy to calculate that:
$$
\dot{\Omega}={\varphi}'\dot{v}+{\varphi}''v^2, \eqno(10b)
$$
$$
\dot{m}=m{\gamma}^2[(\Omega+r\varphi''v)r{\Omega}v+(v+r^2\varphi'
\Omega)\dot{v}]; \eqno(10c)
$$
and using these relations together
with (8) we can easily derive the explicit equation for
the radial acceleration of the bead:
$$
\ddot{r}={{r\omega\Omega-{\gamma}^2rv(\varphi'+\omega v)
(\Omega+r\varphi''v)}\over{\gamma^2\Delta^2}}, \eqno(11)
$$
where
$$
\Delta \equiv [1-{\omega}^2r^2+\Phi^2]^{1/2}. \eqno(12)
$$

The Eq. (11) being of the form $\ddot{r}=G(\dot{r},r)$ admits full
numerical solution, as the standard initial value
problem, providing the initial
position of the bead, $r_0$, its initial velocity, $v_0$, and the
shape of the pipe, $\varphi(r)$, are specified.

Defining the spatial vector of the 2-velocity ${\bf v} \equiv
(v, r \Omega)$, we can calculate the absolute value of the reaction
force $|{\bf F}|$ using the equation [26]:
$$
\dot{m}={\bf F}\cdot{\bf v}, \eqno(13a)
$$
which, in our case, leads to:
$$
\sqrt{g_{rr}}\dot{m}=r \omega |F|. \eqno(13b)
$$

It is also easy to verify that the following quantity:
$$
\Psi \equiv m(t) - \omega r P_{\varphi} = m_0\gamma(1-r^2\omega
\Omega)=const(t), \eqno(14)
$$
is the {\it constant} in time. This
allows to find the solutions of the problem as functions of the
specific value of this constant. In the next subsection we will
see what is the physical meaning of this parameter - it turns out
to be proportional to the proper energy of the moving bead in the RF.

One important class of a possible shape of the curved trajectory
is {\it Archimedes spiral}, given by the formula:
$$
\varphi(r)=ar,~~~~a=const. \eqno(15)
$$
In this case, since
$\varphi''=0$, from (11) it is easy to see that $\ddot{r} \sim
\Omega$, while (14) implies that $|{\bf F}| \sim \Omega$ as well.
Therefore, in the case of the Archimedes spiral trajectory, we can
predict that the asymptotic behavior of the functions $\Omega(t)$,
$v(t)$, $\dot{v}(t)$, and $|{\bf F}(t)|$ will be similar.

\subsubsection{RF treatment}

We see the LF treatment allows to solve the problem and to obtain
the complete information about the dynamics of the bead motion along the
fixed nonstraight (curved) trajectories. However it is
quite instructive and much more convenient to consider the same
problem in the frame of reference, rotating with the pipe-bead
system (rotating frame - RF). In order to do this we, first, need
to switch from (4) to the frame, rotating with the angular velocity
$\omega$. Employing the transformation of variables:
$$
T=t, \eqno(16a)
µ$$
$$
X=rcos \phi = rcos(\varphi + \omega t), \eqno(16b)
$$
$$
Y=rsin \phi = rsin(\varphi + \omega t), \eqno(16c)
$$
we arrive to the metric:
$$
ds^2 \equiv = -(1- \omega^2r^2)dt^2 +
2\omega^2dtd\varphi + r^2d\varphi^2 + dr^2. \eqno(17)
$$

For the straight pipe ($\varphi=\varphi_0$) case (17) reduces
to the metric $ds^2= -(1- \omega^2r^2)dt^2+ dr^2$, which was basic
metric for the [12] study. Now, for a curved pipe, defined
by the equation (1), (17) reduces to the following form:
$$
ds^2 = -(1- \omega^2r^2)dt^2 + 2\omega r\Phi dtdr +
(1 + \Phi^2)dr^2. \eqno(18)
$$

For the resulting metric tensor
$$
\|g_{{\alpha}{\beta}}\|
\equiv
{\left(
        \matrix{-(1-\omega^2r^2), & \omega r\Phi   \cr
                \omega r\Phi,     & 1+\Phi^2       \cr}
\right)}, \eqno(19)
$$
we can easily find out that
$$
\Delta \equiv [-\det(g_{{\alpha}{\beta}})]^{1/2}=(1-{\omega}^2r^2+
{\Phi}^2)^{1/2}, \eqno(20)
$$
and, apparently it is the same function $\Delta$, defined previously
by (13).

For this relatively simple, but nondiagonal, two-dimensional spacetime
we can develop the $``1+1"$ formalism. Doing so we follow as a
blueprint the well-known ``3+1´´ formalism, widely used in the
physics of black holes [27-29]. Namely, we introduce definitions
of  the {\it lapse function}:
$$
\alpha \equiv{{\Delta}\over{\sqrt{g_{rr}}}}=\sqrt{{{1-\omega^2r^2+\Phi^2}
\over{1+\Phi^2}}}, \eqno(21)
$$
and the one-dimensional vector
$\vec{\beta}$ with its only component:
$$
\beta^r\equiv {{g_{tr}}\over{g_{rr}}}={{{\omega}r{\Phi}}\over
{1+\Phi^2}}. \eqno(22)
$$

Within this formalism (18) can be presented in the following way:
$$
ds^2=-\alpha^2dt^2+g_{rr}(dr+\beta^r dt)^2. \eqno(23)
$$

Note that for the metric tensor (19) $t$ is the cyclic coordinate and,
moreover, in the RF the motion of the bead inside the pipe is {\it
geodesic} - there are no external forces acting on it. Hence the
{\it proper energy} of the bead, $E$, must be a
conserved quantity. Employing the definition of the four velocity
$U^{\alpha}{\equiv}dx^{\alpha}/d{\tau}$ we can write:
$$
E \equiv- U_{t}= -U^t{\left[g_{tt} + g_{tr}v\right]}  =
{\it const}. \eqno (24)
$$

On the other hand, the basic four-velocity normalization condition
$g_{{\alpha}{\beta}}U^{\alpha}U^{\beta}=-1$ requires
$$
U^t={\left[-g_{tt}-2g_{tr}v-g_{rr}v^2\right]}^{-1/2}, \eqno(25a)
$$
this equation, written explicitly, has the following form:
$$
U^t={\left[1-\omega^2r^2-2\omega r \Phi v-(1+\Phi^2)v^2\right]}^
{-1/2}. \eqno(25b)
$$
Recalling the expression (3) for the angular velocity of the
bead $\Omega(t)$, measured in the LF, and the definition (5) of the
Lorentz factor $\gamma(t)$ in the same frame of reference we can
easily see that:
$$
U^t=[1-r^2\Omega^2-v^2]^{-1/2}=\gamma(t). \eqno(25c)
$$

It is important to note that the conserved proper energy of the
bead, $E$, defined by (24) may be written in terms of the $\Omega(t)$
and $\gamma(t)$ functions simply as:
$$
E=\gamma(t)[1-r^2(t)\omega\Omega(t)]=const. \eqno(26)
$$
Taking time derivative of this relation and rearranging the terms
we will finally arrive to exactly the same Eq. (11) for the radial
acceleration of the bead $\ddot{r}$ as in the LF treatment. Note also
that ${\Psi}=m_0 E$.

But the convenience of the RF treatment goes much further. From (24)
and (25a) we can derive the explicit quadratic equation for the
velocity:
$$
(g_{tr}^2+E^2g_{rr})v^2+2g_{tr}(g_{tt}+E^2)v+g_{tt}(g_{tt}+E^2)=0,
\eqno(27)
$$
with the obvious solution:
$$
v=\dot{r}={{\sqrt{g_{tt}+E^2}}\over{(g_{tr}^2+E^2g_{rr})}}
{\left[-g_{tr}\sqrt{g_{tt}+E^2} \pm E \Delta \right]}. \eqno(28)
$$

The $``1+1"$ formalism helps to write equivalents of the same
equations in a more elegant form. Namely, if we define the radial
velocity
$$
V^r \equiv {{1}\over{\alpha}}{\left(v+\beta^r\right)},
\eqno(29)
$$
and corresponding Lorentz factor:
$$
\tilde{\gamma}
\equiv (1-V^2)^{-1/2}, \eqno(30)
$$
then, instead of (25c), we will simply have:
$$
U^t = \tilde{\gamma}/\alpha, \eqno(31)
$$
while from the (24) we obtain:
$$
E = \tilde{\gamma}[\alpha - (\vec{\beta}\cdot{\vec V})]. \eqno(32)
$$
Instead of (27) we will have [$V^2 \equiv
g_{rr}V^rV^r=V_rV^r$, $\beta^2 \equiv g_{rr}\beta^r\beta^r=\beta_r
\beta^r$]:
$$
(\beta^2+E^2)V^2-2\alpha({\vec{\beta}}{\cdot}{\vec V})+(\alpha^2-
E^2)=0, \eqno(33)
$$
with the solution:
$$
V^r={{1}\over{\beta^2+E^2}}{\left[\alpha\beta^r \pm E
\sqrt{{{E^2+\beta^2-\alpha^2}\over{g_{rr}}}}~\right]}. \eqno(34)
$$

Note that the RF Lorentz factor defined by (30) and the LF Lorentz
factor, specified by (5) do {\it not} equal each other
$$
\tilde{\gamma}(t) \ne \gamma(t), \eqno(35)
$$
which is the manifestation
of the obvious fact that Lorentz factor is not an invariant physical quantity.
One can see that for the $\tilde{\gamma}(t)$ the following quadratic
equation holds:
$$
(\alpha^2-\beta^2){\tilde{\gamma}}^2-2\alpha E \tilde{\gamma}
+(E^2+\beta^2)=0, \eqno(36)
$$
with the following solution:
$$
\tilde{\gamma}(t)={{1}\over{\alpha^2-\beta^2}}{\left[\alpha E
\pm|\beta|\sqrt{\beta^2+E^2-\alpha^2}~\right]}. \eqno(37)
$$

Therefore, above developed theory allows us to look for the
solution of the initial value problem for the bead moving along an
arbitrarily curved pipe. The thorough consideration of different
particular cases is beyond the scope of the present paper.
Instead, we will give representative solutions for one of the
simplest kinds of a spiral -- Archimedes spiral -- defined by
(15).

The scheme for the complete inspection of the problem for any given
initial value problem is the following: First, one specifies the
initial location of the bead ($r_0$) and its initial radial
velocity ($v_0$). The values of the $\varphi'(r_0)$ and
$\Phi(r_0)$ are fixed as soon as we specify the form
of the pipe $\varphi(r)$. The initial values for the $\Omega(0)$
and $\gamma(0)$ are given by (3) and (5):
$$
\Omega_0=\omega+\varphi'(r_0)v_0, \eqno(38)
$$
$$
\gamma_0=[1-r_0^2\Omega_0^2-v_0^2]^{-1/2}, \eqno(39)
$$
while the
value of the bead proper energy, according to (26), is given as:
$$
E=\gamma_0[1-r_0^2\omega\Omega_0]. \eqno(40)
$$

Working with the  Eq.(28), as the first order ordinary
differential equation for the radial position $r(t)$ of the bead
at any moment of time, we can subsequently calculate all other
physical variables. On the Fig.1 the set of solutions is given for
the case of the Archimedes spiral with $a=-5$, rotating with the
angular velocity $\omega=2$  for the bead, which initially was
situated right over the pivot of the rotator ($r_0=0$) and had
initial radial velocity $v_0=0.1$.

These plots tell us that in the limit of large distance from the
rotator the value of the radial velocity tends to the asymptotic
value:
$$
\liminf{v(t)}=v_\infty \equiv -\omega/a,  \eqno(41)
$$
which, in this case, is equal to $v_\infty = 0.4$.

\begin{figure}
\epsfig{file=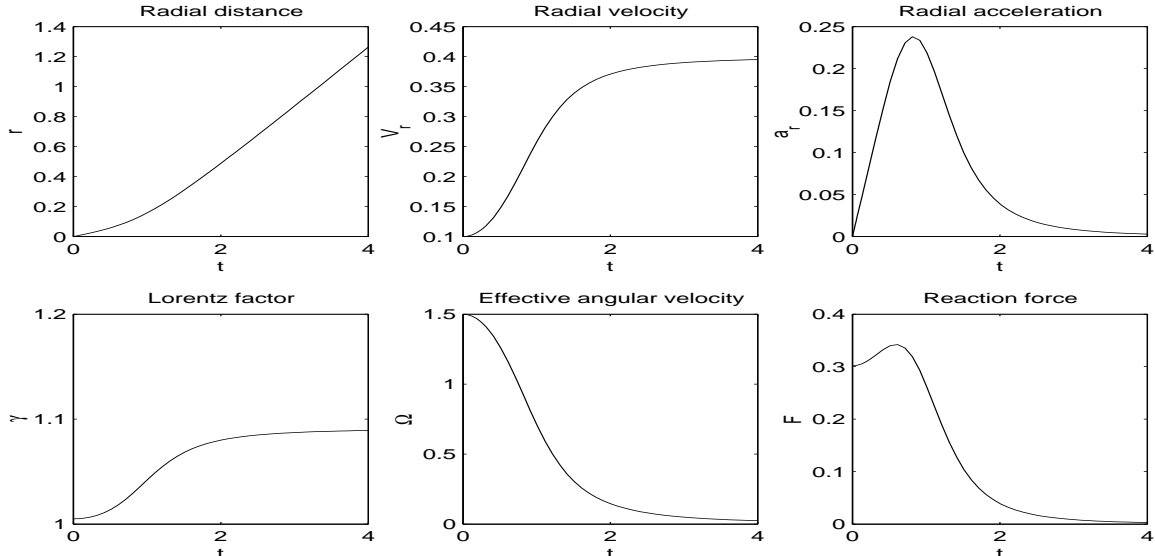, angle=0, width=15.5cm, height=7.5cm}
\caption{Graphs for the radial distance $r(t)$, velocity $v(t)$,
acceleration $a(t)$, the Lorentz factor $\gamma(t)$, angular
velocity $\Omega(t)$ and the absolute value of the reaction force
$|{\bf F}(t)|$ for the rotationally ($\omega=2$) driven bead,
moving on the Archimedes spiral with $a=-5$. $r_0=0$, $v_0=0.1$.}
\end{figure}

The angular velocity of the bead in the LF tends to zero, as well
as the absolute value of the pipe reaction force, implying that at
infinity the bead asymptotically reaches the limit of the
force-free motion. This limit is understandable also analytically,
because from (3) and (28) we can see that:
$$
v\to\frac{\omega}{|a|}+\frac{E\sqrt{a^2-\omega^2}}{\omega a^2
r^2}, \eqno(42)
$$
$$ \Omega\to \frac{E\sqrt{a^2-\omega^2}}{\omega
a r^2}. \eqno(43)
$$

From these expressions it is clear that this regime is accessible
iff the condition $|a|>\omega$ holds! Otherwise, the particle is not
able to reach the infinity.

Since the shape of the function $r(t)$ is almost linear it is
instructive to make plots for the functions $v(r)$ for
different values of the initial radial velocity $v_0$, but with
all other parameters of the initial value problem being the same.
On the Fig.2 we plotted these functions for eight different values
of the initial radial velocity. We see that when $v_0=v_\infty$
the movement of the particle is force-free (geodesic) and uniform
during the whole course of the motion. Physically it means that for
this particular value of the $v_0$ the shape of the pipe follows the
geodesic trajectory of the
bead, in the RF, for the metric (17) on the rotating 2D disk, so
the bead
moves freely, without interacting with the walls of the pipe.
When $v_0<v_\infty$, the
particle moves with positive acceleration and asymptotically
reaches the force-free regime in the infinity. While, when
$v_0>v_\infty$ the character of the motion is decelerative, but
the force-free limit is reached, again, when the bead heads to
infinity.

One more example of the latter behavior, similar to the case shown
on the Fig. 1, but plotted for the initial
velocity $v_0=0.5>v_\infty=0.4$ is given on
the Fig.3. Here we see that, unlike the case given on the Fig.1,
the acceleration of the bead is negative all the time and it
reaches zero ``from below", taking  less and less negative
values. While the angular velocity of the bead relative to the
LF $\Omega(t)$ is also negative from the beginning but its absolute
value decreases and reaches the zero as the particle tends to
the infinity.

\begin{figure}
\epsfig{file=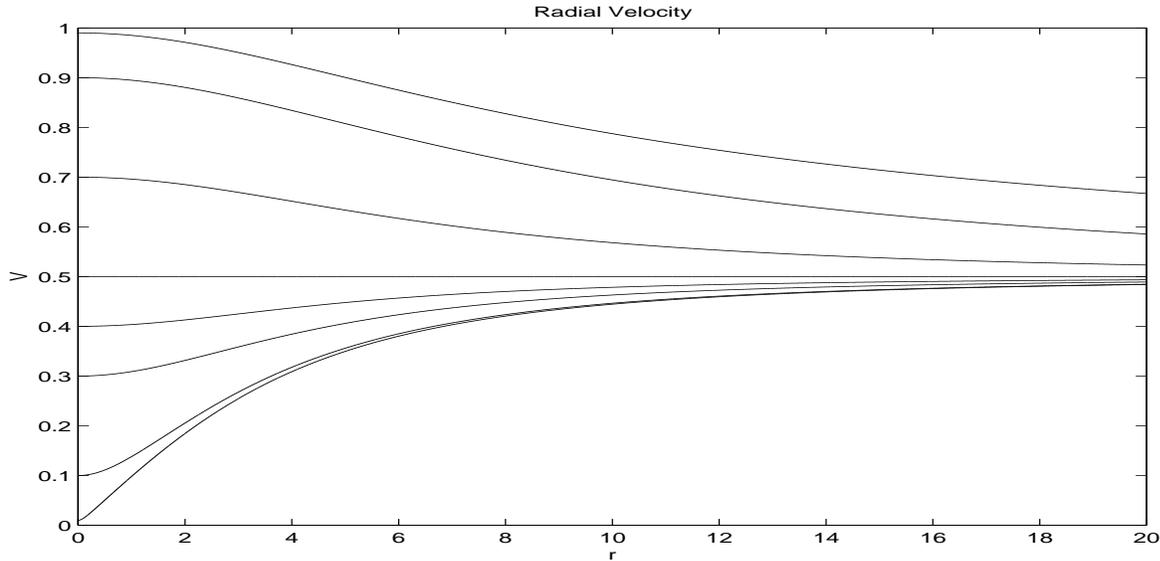, angle=0, width=15.5cm, height=7.5cm}
\caption{Graphs for the radial velocity $v(r)$, when the initial
value of the $v(r)$ is taken to be: $0.01$, $0.1$, $0.3$, $0.4$,
$0.5$ (force-free value), $0.7$, $0.9$, $0.99$. $\omega=0.1$,
$a=-0.2$.}
\end{figure}

\begin{figure}
\epsfig{file=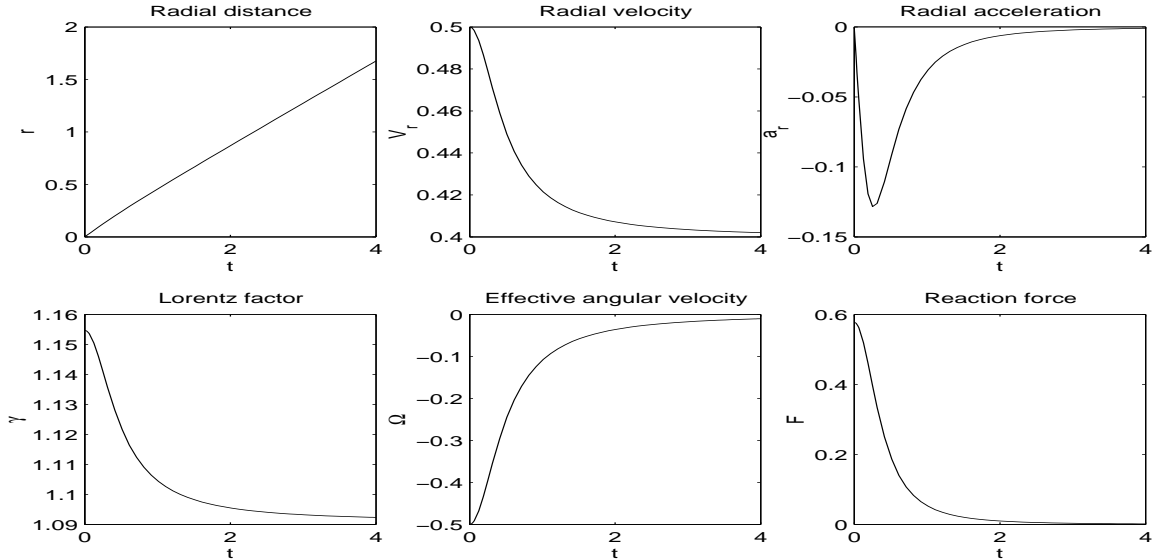, angle=0, width=15.5cm, height=7.5cm}
\caption{Graphs for the radial distance $r(t)$, velocity $v(t)$,
acceleration $a(t)$, the Lorentz factor $\gamma(t)$, angular
velocity $\Omega(t)$ and the absolute value of the reaction force
$|{\bf F}(t)|$ for the rotationally ($\omega=2$) driven bead,
moving on the Archimedes spiral with $a=-5$. $r_0=0$, $v_0=0.5$.}
\end{figure}

\subsection{Conserved energy case}

When the bead accelerates, it continuously takes energy
from the rotator. So,
if one needs to keep the rotation rate constant, one needs
to supply the system with energy from outside. This is, certainly,
less realistic setup than the assumption that the ``rotator-pipe-bead"
system
is conservative, viz. its total energy $E_{tot}$ is constant. In
this case, however, the bead acceleration can not be permanent, because
asymptotically it extracts all energy from the rotator and
reaches the regime:
$\omega(t)\to 0$, $E_M \to 0$ and $E_m \to E_{tot}$. Clearly,
in this situation, it is more convenient to study the
dynamics in the laboratory frame of reference (LF),
in which the rotator and the pipe are rotating rigidly
with the time-dependent angular velocity $\omega(t)$. As regards
the bead, since the shape of the pipe is curved, its angular velocity
relative to the LF is given
by (3). Since the pipe is considered to be massless and absolutely
rigid, it does not contribute any amount of energy and/or
angular momentum to the total energy $E_{tot}$ and angular momentum
$L_{tot}$
of the system. The rotator for simplicity is assumed to be a sphere
of the radius $R$ and the mass $M$ having the inertia moment
$$
{\cal I} = {2 \over 5}MR^2, \eqno(44)
$$
the energy
$$
E_M={{\cal I} \over 2}\omega^2(t),\eqno(45)
$$
and the angular momentum
$$
L_M={\cal I} \omega(t). \eqno(46)
$$

Note that (44-46) are nonrelativistic
expressions. If initially, at $t=t_0$,
$$
\omega_0R \ll 1, \eqno(47)
$$
then it will remain nonrelativistic during the whole
course of the motion, because it is assumed that the bead
constantly extracts energy from the rotator, while the latter slows
down so that the angular velocity $\omega(t)$ is a monotonically
decreasing function of time. The (47) condition seems to be valid
for the known fastest rotators in the Nature --- pulsars. For the
Crab pulsar, for instance, $R=1.2\times10^6cm$, $\omega_0=190.4Hz$
and consequently $\omega_0R/c\simeq7.6\times10^{-3}$. Even for the
fastest millisecond pulsars $\omega_0R/c\le0.25$. This justifies
the usage of nonrelativistic (44-46) expressions in our analysis.

The remaining part of the threefold system --- the bead --- is
assumed to be of the rest mass $m_0$. Its angular velocity and
radial velocity relative to the LF, at any given moment of time, are
$\Omega(t)$ and $v(t) \equiv \dot{r}$, respectively. Even when the
initial radial velocity of the bead is nonrelativistic ($v_0 \ll
1$), it is still necessary to write relativistic expressions for
its energy and angular momentum, because
the bead gains energy, accelerates and sooner or later its motion
becomes relativistic. Therefore, its energy and angular momentum
must be written as:
$$
E_m=m(t)=m_0\gamma(t), \eqno(48)
$$
$$
L_m=m_0\gamma(t)r^2(t)\Omega(t)=m(t)r^2(t)\Omega(t), \eqno(49)
$$
where the LF Lorentz factor $\gamma(t)$ is defined by (5).

The system ``rotator-pipe-bead" is conservative, there is no
energy inflow from outside. In this sense it principally differs
from the one considered in the previous section, where either the
external energy source was necessary to keep the rotation rate
$\omega$ constant or the rotator was assumed to possess an infinite
amount of energy. Now, since the system is conservative, its
dynamics are governed by the conservation laws of its total energy
$E_{tot} \equiv E_M+E_m$ and total angular momentum $L_{tot}
\equiv L_M+L_m$:
$$
{{{\cal I} \over 2}}~\omega^2(t)+m_0\gamma(t)=E_{tot}, \eqno(50)
$$
$$
{\cal I} \omega(t)+m_0\gamma(t)r^2\Omega(t)=L_{tot}. \eqno(51)
$$

And the solution of the problem reduces to the solution of these
equations, linked with (9), for two unknown
functions of time $r(t)$ and $\omega(t)$ for an arbitrary initial
value problem: initial location of the
bead $r_0=r(0)$ and the initial value of the rotation rate
$\omega_0=\omega(0)$ of the whole system. The detailed study of this problem
is beyond the scope of this paper and will be given in a separate publication.

\section{Conclusion}

The purpose of the present paper was to study the dynamics of
relativistic rotating particles with prescribed, curved trajectories
of motion in the rotating frame of reference. The
work is a natural generalization of the {\it gedanken}
``pipe-bead" experiment considered Machabeli
and Rogava [12]. In that paper the authors considered the case of
the {\it straight} rotating pipe and they found out that when the
velocity of the bead, driven by the rotation of the whole device
and sliding along/within the pipe, is high enough the character of
the motion changes from the accelerated to the decelerated one. In
particular, it was found that when the bead starts moving from the
pivot ($r=0$) of the rotating pipe with initial velocity $v_0 >
\sqrt{2}/2$, the motion is decelerative from the very beginning.

In this paper we consider the motion of rotationally driven particles
along flat
trajectories of arbitrarily curved shape. The practical motivation
for this approach and its importance are related with the following
two facts:

\begin{enumerate}

\item

The `pipe-bead' (or the `bead-on-the-wire') {\it gedanken} experiment
is considered as a model for the study of dynamics of
centrifugally driven relativistic particles in rotating
magnetospheres, in various classes of astrophysical objects, like
pulsars [17-18,12,20] and AGNs [21-24]. The role of ``pipes" is
played by the magnetic field lines.

\item

The shape of magnetic field lines is always curved. It implies that for the
large-scale, global dynamics of charged particles ---
driven by centrifugal forces and moving along curved field
lines of rotating magnetospheres --- it is important to
know what qualitative changes occur when the form of the field lines
is not linear but curved.

\end{enumerate}

In this paper we studied this problem, on the level of the idealized
{\it gedanken} experiment, both in the laboratory (LF) and in the
rotating (RF) frames of reference. For the simple example of the
Archimedes spiral we found that the dynamics of such particles
may involve both accelerative and decelerative modes of motion.

One important difference from the linear pipe case [12] is that
for the case of a curved pipe the motion of the bead is not any
more radially bounded: there exist regimes of motion when the bead
may reach infinity. This result has simple physical explanation.
For the case of the linear pipe, rotating with the constant
angular velocity $\omega_0$, the natural limit of the radial
motion was given by the light cylinder radius, defined as $R_L
\equiv \omega_0^{-1}$. Now, in the case of the curved pipe, even
when it rotates with the same constant rate, the bead can slide in
the azimuthal direction, following the curvature of the pipe and
having a variable angular velocity ${\Omega}(t)$. It means that
now the role of the {\it effective light cylinder} is played by $R_L(t)=
{\Omega}(t)^{-1}$, and, hence, all those radial distances become accessible,
where $r(t)<R_L(t)$. Therefore, if both $r(t)$ and $R_L(t)$ are
monotonically increasing functions, but the former stays always
smaller than the latter (evidently it was the case in above considered examples
for the Archimedes spiral) then the bead can reach infinity.

Moreover, we found that there are special solutions, which are
force-free during the whole course of the motion. These are simply
geodesics in the two-dimensional rotating metric (17). In the LF
the motion of the bead in this case is radial, because its angular
velocity ${\Omega}(t)$ stays zero all the time and,
correspondingly, the light cylinder is at the infinity from the
very beginning. The form of the trajectory in the RF in this
case, $\varphi(r)=-(v_0/\omega_0)r$, is simply that trace, which a
free bead could leave on the surface of the rotating disk during
the course of its geodesic motion. Intuitively it is evident that if the
pipe has this particular form the bead slides within it freely,
without interaction with the walls of the pipe.

We considered only one, simple, subclass of spiral trajectories as
the representative example of the solutions, but the developed
theory may be used for the study of the dynamics of particles
moving along arbitrarily shaped flat trajectories. It means that
this approach may find wide applications to different astrophysical
situations where rotation impels plasma particles to move along
curved magnetic field lines.

We also gave basic equations and outlined the scheme for the solution
of the more general version of the
same problem, where the angular velocity of the rotating system is
not assumed to be constant. Instead, it is assumed that the system
rotator-pipe-bead is {\it conservative} and the rotator is allowed
to exchange perceptible portions of energy with the bead.

The formalism developed and the results found in this paper suggest
some important directions
of the further research, which could
be physically and astrophysically relevant.

First, it seems worthy to consider the case of the charged bead,
which naturally would radiate while performing its nonuniform motion
along the curved pipe. The radiative energy losses and the change
of the angular momentum of the bead due to radiation could affect
the dynamics of the bead and the whole system, considered, again, to
be conservative. In the astrophysical context it could be interesting
to see how the radiation of the bead would appear for the
distant observer.

Second, it is quite natural to try to extend the analysis for the
3-D fixed trajectories of motion, considering a family of
axisymmetric trajectory lines and imitating the structure of the
pulsar dipole magnetic field. This could bring us at least one
step closer to the understanding of the radiative processes in
pulsar magnetospheres.

Third, sooner or later, we should address the fluid (plasma)
problem and try to see how a continuous stream of fluid particles
would behave, moving along rotating curved trajectories. This will
comprise one more step closer to the reality of the pulsar environment
and could help to relate with each other total energy losses of a
pulsar, estimated through its slowing rate, and its radiation
losses and energy taken away by centrifugally driven plasma,
forming eventually the pulsar wind.

These efforts would, hopefully, bring us to the construction of
the unified theory of the pulsar magnetosphere, where the inertial
aspects of the particle dynamics would be taken into due account.
One could then try to test the theory with the existing empirical
(observational) data
about the energy deposited by pulsars into  their winds and the
energy they lose via their radio emission. This way we could have
a clue as of how important inertial processes (often unfairly neglected) are
in the dynamics of pulsars.

Similar problems can be addressed also in the context of the
centrifugal acceleration of particles in the jets in AGNs [21-24].

Yet another promising field of application is the accretion of
plasma on strongly magnetized neutron stars, which is believed to
lead to the appearance of {\it X-ray pulsars}. The final stage of
the accretion is dominated by the dipole magnetic field of the
accreting neutron star. It is normally assumed that the motion of
accreted plasma particles is guided to the magnetic poles of the
star by the rotating family of polar magnetic field lines. So, in
a certain sense, this problem is of the same kinematic nature as
the one related with radio pulsars, except that this time plasma
moves from the region outside of the light cylinder towards the
star. Certainly here, again, radiative effects and collective
plasma effects are essential, so the simple 'one-particle´
treatment can give only very approximate picture of the involved
physical processes. But taking into account the plasma fluid
effects and the role of the radiation on the dynamics of infalling
plasma streams, one could try to show  how important the
rotational (inertial) processes are for the dynamics of the flows
infalling on strongly magnetized neutron stars and what is the
influence of these processes on the observational appearance of
related X-ray sources.

\section{Acknowledgements}

The authors are grateful to George Machabeli and Swadesh Mahajan
for valuable discussions. A.D. and Z.O. are grateful to the Abdus
Salam International Centre for Theoretical Physics for the
hospitality and support during their visits to the Centre as a
Regular Associate and a Young Collaborator, respectively.

\end{document}